# A Marching Cube Algorithm Based on Edge Growth

Xin WANG[1], Su GAO[2], Monan WANG [1]*, Zhenghua DUAN[1]

1. Key Laboratory of Medical Biomechanics and Materials of Heilongjiang Province, Harbin University of Science and Technology, Harbin 15000, China
2. Beijing Normal University Hospital, Beijing 100875, China

* Corresponding author, mnwang@hrbust.edu.cn

Received:          Accepted:

Supported by NSFC(No.61972117) and the natural science foundation of Heilongjiang Province of China (No.ZD2019E007)

**Abstract** Marching Cube algorithm is currently one of the most popular 3D reconstruction surface rendering algorithms. It forms cube voxels through the input image, and then uses 15 basic topological configurations to extract the iso-surfaces in the voxels. It processes each cube voxel in a traversal manner, but it does not consider the relationship between iso-surfaces in adjacent cubes. Due to ambiguity, the final reconstructed model may have holes. We propose a Marching Cube algorithm based on edge growth. The algorithm first extracts seed triangles, then grows the seed triangles and reconstructs the entire 3D model. According to the position of the growth edge, we propose 17 topological configurations with iso-surfaces. From the reconstruction results, the algorithm can reconstruct the 3D model well. When only the main contour of the 3D model needs to be organized, the algorithm performs well. In addition, when there are multiple scattered parts in the data, the algorithm can extract only the 3D contours of the parts connected to the seed by setting the region selected by the seed.

## 1. Introduction

3D reconstruction algorithms are mainly divided into volume rendering and surface rendering. The surface rendering algorithm first extracts features through two-dimensional tomographic images, and then uses intermediate geometric primitives such as quadrilaterals, triangles, hexahedrons, cones, and tetrahedrons to fit the 3D surface contour of the model to be reconstructed [1]. The volume rendering algorithm directly maps the input data into a 3D figure on the screen without passing through the intermediate geometric primitives. The research in this article belongs to a surface rendering algorithm. Generally speaking, compared with the volume rendering algorithm, the operation of the surface rendering algorithm requires less memory and performs better in the details of the reconstruction model.

Currently, the main algorithms for surface rendering include: Triangulation of Contour Lines, Cuberille, Dividing Cubes, and Marching Cubes, etc[2-5]. Among them, the MC algorithm was proposed by Lorensen and Cline in 1987. Because of its good 3D reconstruction effect, fast reconstruction speed and simple algorithm principle, it has won the recognition of the majority of image developers in practical applications. Nowadays, MC algorithm has become a classical algorithm in the field of 3D reconstruction. With people's in-depth research and application, we found that the MC algorithm has the following shortcomings: The first point is ambiguity, which means that the topological configuration between adjacent cubes does not match, and the finally reconstructed 3D model exists holes. The second point is that some time is wasted on the calculation of empty voxels. An empty voxel refers to a cube that does not intersect with the iso-surface, generally accounting for more than 95% of all cubes[6]. Because the MC algorithm processes all cube voxels in a traversal manner, part of the time is wasted on the calculation of empty voxels. The third point is that the MC algorithm will generate a large number of triangles. Too many triangles are not conducive to the rendering and storage of the model.

The ambiguity of the MC algorithm was originally proposed by Dürst et al in 1988[7]. They found that the MC algorithm has some basic configurations that can be replaced by a variety of other topological configurations, and it is impossible to determine which topology to use during reconstruction. The irrational selection of topological configuration may lead to the lack of connection between the iso-surfaces in the adjacent voxels, and finally lead to the existence of holes in the generated iso-surfaces. The Marching Tetrahedra (MT) algorithm proposed in 1990[8], as one of the earliest surface rendering algorithms to solve ambiguity, MT is based on the MC algorithm, which subdivides each cube voxel into multiple four sides voxels, and then extract iso-surfaces from each tetrahedron for reconstruction, but this method also has ambiguity when the cube is divided into tetrahedrons. Due to the ambiguity of segmentation, the reconstructed model may still have holes. In addition, MT will be much longer than MC algorithm in reconstruction time. In 1991, Nielson proposed the Asymptotic Decider[9]. This method can better judge the ambiguity of the cube surface. According to the criterion of judgment, the ambiguity of the cube surface can be resolved, but the method still cannot resolve the ambiguity of the cube. Regarding the ambiguity of MC algorithm body, the current main judgment methods include saddle point method, critical point method and generalized asymptotic method. For these judgment methods, although the ambiguity is resolved to a certain extent, it also increases the time for reconstruction. In 1990, Wilhelms proposed that only 3% of topological structures are ambiguous in practical applications[10]. Because of the low frequency of ambiguity, Wilhelms proposed an extended look-up table method, which extends the main ambiguous topological structures, the judgment of topological configuration is mainly based on the gray scale and threshold value of cube vertices. This kind of method is of special research value because it not only reconstructs quickly but also reconstructs well under the premise of solving the holes in the model. In 1994, Zhou further improved this method[11]. In 2013, Masala expanded the 15 topological configurations of MC algorithm to 21 based on the extended look-up table method. This method can effectively solve the ambiguity problem, it has good performance in the speed of reconstruction and the quality of 3D model. At present, some software has been applied to this method[12,13].

In the optimization of the reconstruction time of MC algorithm, the current methods include parallel computation, octree and so on. Octree method can avoid the calculation of some empty voxels to some extent by block processing the input data, which was proposed by Wilhelms and Van Gelder[14]. Montani proposed to use the midpoint to replace the linear interpolation point in the MC algorithm. This method reduces a lot of floating-point operations and is a commonly used acceleration method for the current MC algorithm[15]. Based on the Montani's idea, some scholars also use golden ratio points and multi-segment points to replace Linear interpolation points. When the input data is large, the MC algorithm will generate a large number of triangular meshes. The main solution is to simplify the reconstructed 3D mesh model[16]. In 2010, the Simplified Marching Cubes proposed by Vignolet, instead of interpolating the intersection point between the iso-surface and the cube, use the vertex of the cube instead of interpolating the intersection point. This method can effectively avoid the ambiguity of MC algorithm, the number of triangles is less than that of MC algorithm, but the precision of SMC algorithm is not as good as that of MC algorithm[17].

In this paper, a surface rendering algorithm based on edge-growing is proposed to solve the problems of the ambiguity of topology and the slow reconstruction time of MC algorithm. In this algorithm, seed triangles are selected, and then the edges of the seed triangles are added to achieve the goal of 3D reconstruction.

## 2. Algorithm description

### 2.1 Method and process

In the process of 3D reconstruction of MC, the iso-surface of each cube is interpolated according to its 15 basic topologies, as shown in figure 1.

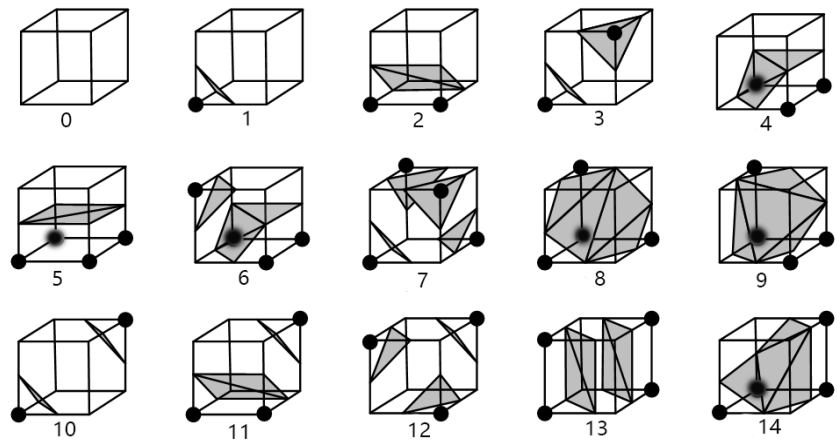

Fig.1 The 15 basic topological configurations of MC algorithm

The ambiguity of the MC algorithm means that the topological configuration of the cube is not unique when selected, and the corresponding topological configurations can be replaced by other topological configurations. Due to the existence of ambiguity, and the MC algorithm processes each cube separately in the reconstruction process, it does not consider the relationship between adjacent cubes. When not sure which topology to use, sometimes the reconstructed model will have holes, as shown in figure 2.

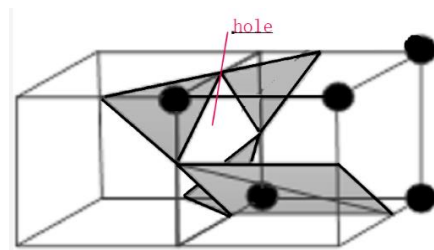

Fig.2 The hole caused by ambiguity

As we all know, for an object, its outer contour is a continuous surface. In 3D reconstruction, the iso-surface of the object is also continuous. Therefore, the iso-surfaces left by passing through adjacent cubes do not exist independently. They are connected to each other, as shown in figure 3.

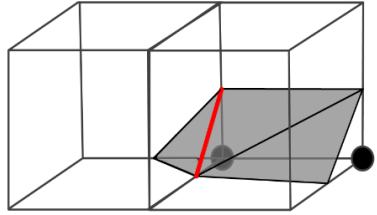

**Fig.3 The relationship between the iso-surfaces of adjacent cubes**

As can be seen from figure 3, when an iso-surface passes through a pair of adjacent cubes, the iso-surface leaves a line of intersection on the common surface of the adjacent cubes, and this line of intersection is the common edge of the iso-surface in the two cubes, it acts as a connection between the two iso-surface. Based on the characteristic of intersecting lines, this paper presents an edge-growing MC algorithm. The core of the algorithm is to use the existing iso-surface to generate the iso-surface in the adjacent cube. As shown in figure 3. If the iso-surface in the left cube has been generated, but the right one does not yet exist, then the iso-surface in the right cube can be generated through the common edge of the iso-surface in the two cubes, for this common edge is called the growth edge.

The algorithm presented in this paper has a growth queue and two labeled arrays to keep the whole reconstruction process running smoothly. The queue stores the growth edge information, the tag arrays record how each cube is processed and how the cube's growth is put into the growth queue. The growing edge information includes the 3D coordinates of the two ends, the gradient values of the two ends, and the cube coordinates to which the growing edge is to be grown. The algorithm is divided into the following steps:

Step 1: Select the seed triangles, and put the growth edge formed by the seed triangles into the growth queue.

Step 2: Take out the growth edge from the growth queue, interpolate the intersection points of the iso-surface and the cube edge according to the topology configuration, and connect the intersection points to form a triangular mesh.

Step 3: Put the new edge where the iso-surface generated in Step 2 intersects with the cube surface as the growth edge and put it in the growth queue.

Step 4: Repeat steps 2 and 3 until the queue is empty.

The flowchart is shown in figure 4.

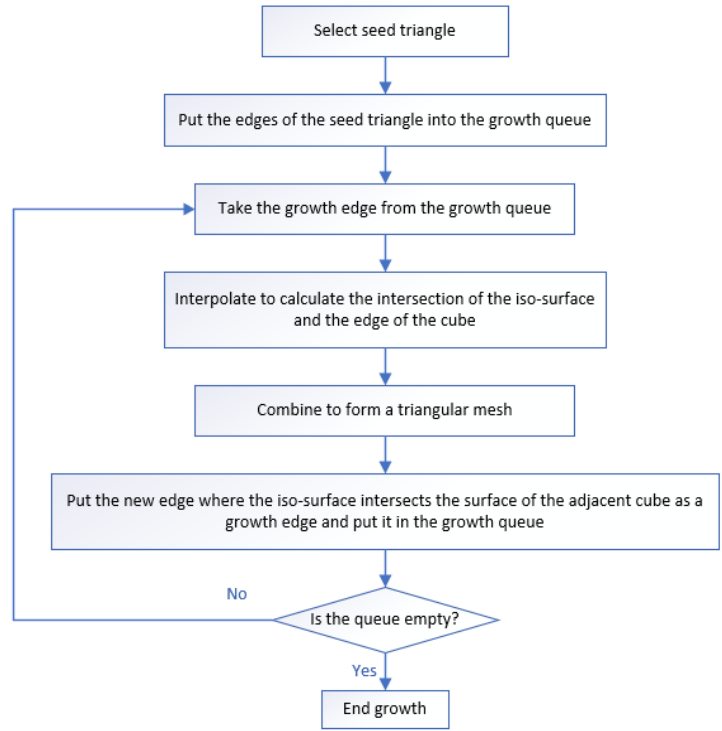

**Fig.4 The algorithm flowchart proposed in this paper**

## 2.2 Seed triangle selection

As can be seen from figure 1, the iso-surface corresponding to topology configuration 1 of MC algorithm is a triangle. In view of the simplicity of topological configuration 1, three sets of data are used for 3D reconstruction. The results of 3D reconstruction are shown in figure 4. Figures (a), (b) and (c) are the results of 3D reconstruction of data 1, 2 and 3 using MC algorithm respectively, figures (d), (e), and (f) show the effect of data 1, 2, and 3 reconstruction when only the cube of topological configuration 1 is reconstructed. From the comparison of the results of the same data reconstruction, figures (d), (e) and (f) can basically reflect the approximate outline of the object, except for the small number of triangular patches in the model, and the independent triangular patches are evenly distributed on each cross-section. From the results of reconstruction, we can see that topological configuration 1 has a high proportion in the 3D reconstruction of these models.

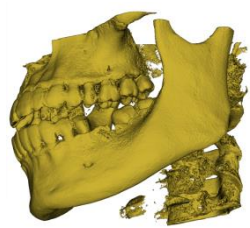
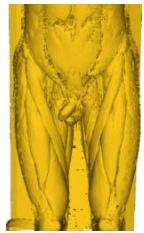
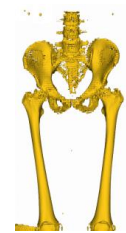

(a)  (b)  (c)

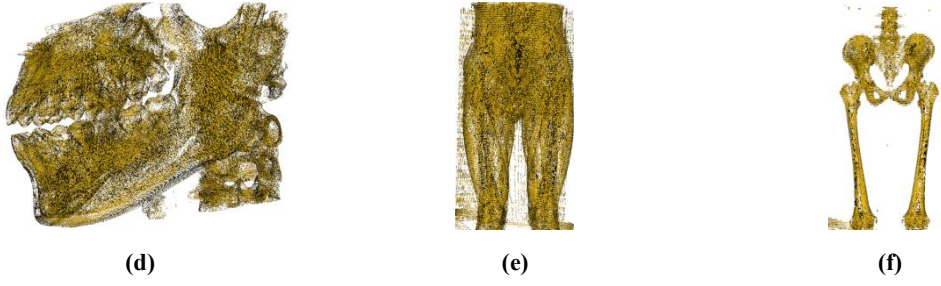

|  (d)  |  (e)  |  (f)  |

**Fig.5 Comparison of the reconstruction results of the MC algorithm and only using topology configuration 1**

Based on the simplicity and high proportion of topological configuration 1, the triangle of topological configuration 1 is used as seed triangle, and then the edge of seed triangle is used as growth edge for growth reconstruction.

In practical application, the number of selected seed triangles must be reasonable, too little may not reach the effect of reconstruction because of noise, too much will spend a lot of time on the selection of seed triangles. In most of the experiments in this paper, all the triangles generated from one layer topology 1 are selected as seed triangles. This method is simple and easy to implement, and it can achieve a good reconstruction effect. For the selection of the number of layers, we need to select the layer whose curvature changes, because the content of topological configuration 1 in this layer is more, the middle layer can usually meet the requirement of seed selection. For the seed triangle, it is not limited to a layer of cubes, but can be selected manually or in a designated 3D area. The core of this algorithm is 3D reconstruction based on edge growth. The purpose of selecting the seed triangle is to get the growing edge, which can also be obtained from other topologies.

### 2.3 3D reconstruction based on edge growth
#### 2.3.1 Interpolation

In this algorithm, a three-segment interpolation method is proposed for interpolation reconstruction. The three-segment interpolation is an interpolation between linear interpolation and midpoint selection.

Assuming that $p_1$ and $p_2$ are the 3D coordinates of two vertices on one side of the cube, $hu_1$ and $hu_2$ are the gray values of the two vertices, and the threshold of the iso-surface is $Y$. When this edge and the iso-surface have the intersection point, the intersection point coordinates are calculated using the following formula.

$$p = p_1 + (p_2 - p_1)(Y - hu_1)/(hu_2 - hu_1) \tag{1}$$

If the $(Y - hu_1)/(hu_2 - hu_1)$ in the formula is called the interpolation ratio $k$, then:

$$p = p_1 + k(p_2 - p_1) \tag{2}$$

The selection of $k$ value in the expression is the special feature of the three-segment interpolation method in this paper. Firstly, the lower limit $m$, the upper limit $n$, the lower value $q$ and the upper value $p$ should be set, in which $0 < q < m < n < p < 1$ and $q + p = 1$ are required. Then, the absolute value of $k$ is compared with $m$ and $n$. When the absolute value of $k$ is less than $m$, the absolute value of $k$ is $q$. When the absolute value of $k$ is greater than $n$, the absolute value of $k$ is $p$. In other cases, the absolute value of $k$ is 0.5. In these three methods, the positive and negative requirements of $k$ are always the same as before, only the absolute value of the value is normalized. In the practical application of this paper, $q$ is 0.25, $m$ is 0.3, $n$ is 0.7, and $p$ is 0.75.

### 2.3.2 topological configurations

Because this paper adopts three-segment interpolation in interpolation criterion, the intersection points of interpolation will only be between two endpoints on each side.

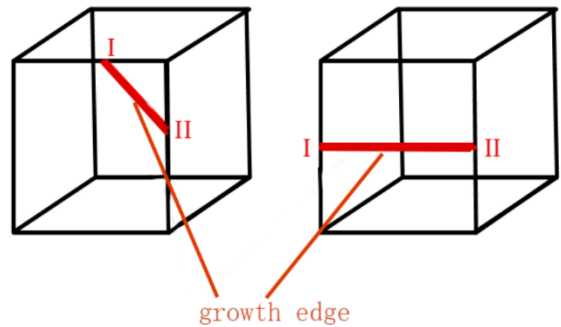

Fig.6 Two kinds of growth edges

The growth edge can be divided into two types according to the position of the two endpoints. The first is that the two ends of the growing edge are on the two adjacent edges, and the growing edge intersects the two sides to form a triangle, as shown on the left side of figure 6. The other is that the two ends of the growing side are on opposite sides, and the growing side divides the surface of the cube into two quadrangles, as shown on the right side of figure 6. In this paper, the former growth edge is called triangle edge, and the latter is called quadrangle edge.

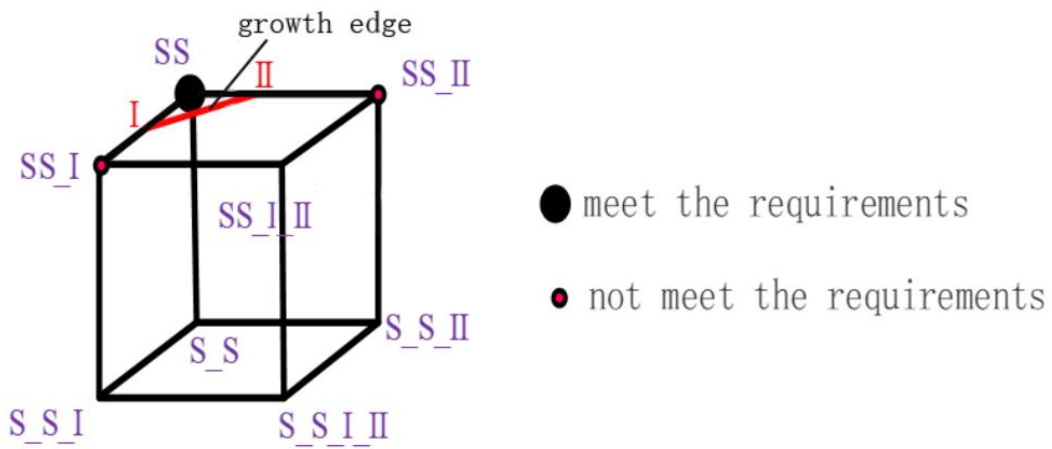

Fig.7 Name of each vertex of the cube when growing with triangle edge

Since the position of the growth edge is uncertain, we name the vertices of the cube based on the two endpoints $I$ and $II$ of the growth-edge and the plane where the growth edge is located, which is conducive to the extraction of the iso-surface. Figure 7 shows the names of the vertices of the cube when the triangle edges are growing. The naming rule is: the names of all vertices on the surface where the growth edge is located begin with SS, the surface where the growth edge is located is called the SS surface, the names of the vertices opposite to the SS all begin with S_S, and the surface they are on is called the S_S surface. The vertices that just form a triangle on the SS surface and the growing edge are directly named SS, and the two vertices adjacent to the point SS on the SS surface are named SS_I and SS_II. The diagonal point opposite to the SS point on the SS surface is named SS_I_II.

The name of each point on the S_S surface mainly refers to the name of the point on the adjacent SS surface. The names of the adjacent points on the two surfaces are the same in the suffix.

When using the triangular edge for growth 3D reconstruction, take the point SS as the reference point, and compare its gray value $Hu_{ss}$ with the reconstruction threshold $Y$. If $Hu_{ss}$ is greater than threshold $Y$, the vertex with gray value greater than threshold $Y$ is said to meet the requirements, and the remaining points are points that do not meet the requirements. $Hu_{ss}$ is not greater than threshold $Y$, then the vertex with gray value not greater than $Y$ is called the points that meet the requirements, the remaining points are the points that do not meet the requirements. As shown in Figure 7, the vertices that meet the requirements are marked with large dots, and the points that do not meet the requirements are marked with small dots. In the subsequent figures, large and small points are also used to mark each vertex. For some unmarked points in some topological configurations, they are called irrelevant points, and they do not participate in the specific iso-surface generation. In addition, because the algorithm in this paper starts from the growing edge, only the iso-surfaces that are connected to the growing edge are extracted in the cube body, so all the points that meet the requirements are connected to each other when extracting the iso-surface.

According to the position of the cube surface where the growth edge is located, and the relationship between the gray value of each vertex of the cube and the threshold value, 11 topological configurations when the triangle edge is used for growth are summarized, as shown in figure 8.

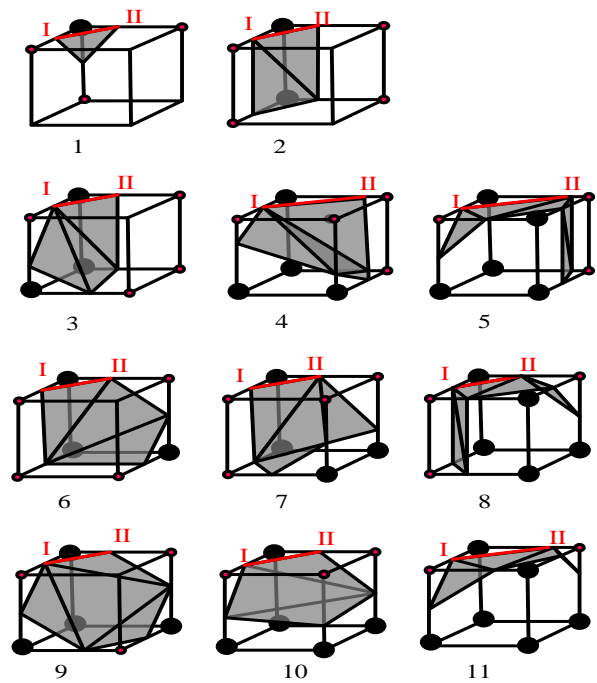

**Fig.8 The topological configurations when growing with triangle edge**

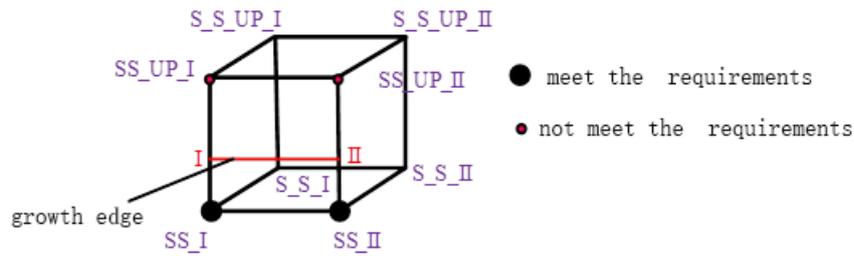

**Fig.9 Name of each vertex of the cube when growing with quadrangle edge**

When growing with four corners, you need to first compare the gray values of the 8 vertices of the cube with the threshold $Y$. And then the vertex is divided into two parts, one is the gray value is greater than the threshold, the other is the gray value is not greater than the threshold. Select the smaller number of points as the reference point, that is, the points that meet the requirements, and the number of points that meet the requirements will be in the closed interval of 1 and 4. As with the growth of triangular edges, when the four corners are used for growth, the vertices of the cube also need to be named. The naming rule is that all the vertices on the surface where the growth edge is located begin with $SS$, and the surface where the growth edge is located is called $SS$ face. The face of the $SS$ face-to-face is the $S\_S$ face, and the names of the vertices on this face start with $S\_S$. On the $SS$ face, the vertex that meets the requirements adjacent to vertex $I$ of the growth edge is called the $SS\_I$ point, and the adjacent points that do not meet the requirements are called $SS\_UP\_I$. The vertex that meets the requirements adjacent to vertex $II$ of the growth edge is called point $SS\_II$ and the adjacent point that does not meet the requirements is called $SS\_UP\_II$. The vertices on side $S\_S$ have the same suffix as the vertices on side $SS$. Figure 9 shows the specific naming of each vertex.

According to the position of the cube surface where the growing edges are located, six topological configurations are concluded when growing with quadrangle edge are used for reconstruction, as shown in figure 10.

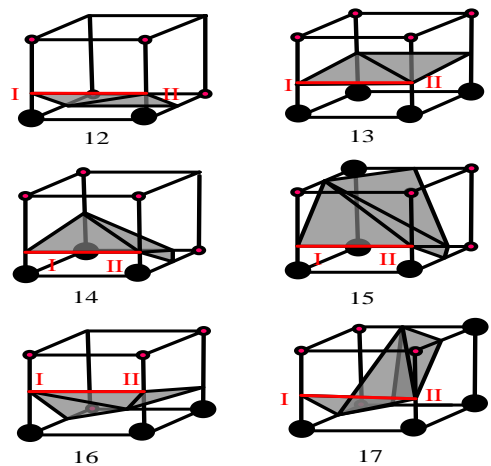

**Fig.10 The topological configurations when growing with quadrangle edge**

The algorithm in this paper extracts the iso-surfaces in the cube according to these 17 topological configurations. After interpolating the iso-surface in the cube according to the topological configuration, it is also necessary to determine whether the intersection of the iso-surface and the surface of the cube can become a new growth edge and continue to grow. The basis for the judgment is that this edge is a new growth edge, and the processing label value and the growth label value of adjacent cubes that share the side are both not 4.

**2.3.3 Cube Marks**

When a cube is processed, its processing needs to be recorded, and only when the processing mark value reaches a certain requirement, then no further processing of the cube is needed. Each cube has two markers, one for processing, which records how the cube was processed according to the growth criteria. Another marker is the growth marker, which records how the cube's growth information is put into the growth queue. There are also five levels for these marks, which are represented by the numbers 0,1,2,3, and 4. Level 0 is the lowest, indicating that the cube has not been processed. Level 4 is the highest, indicating that the cube has been processed and no further processing is required. Other numbers indicate that the cube has been processed but may still need to be processed. The previous 17 topologies show only one growing edge, but in practice, it is possible to have separate iso-surfaces in a cube, so there's a lot of work to be done with cubes like this.

The specific marker assignment criteria are as follows: Extract the cube corresponding to the seed triangle, directly set all the tags of the corresponding cube to 4 after the seed triangle is extracted; After putting the growth edge information into the growth queue, you need to add 1 to the value of the cube growth marker to be processed; When the corresponding topological configuration of the cube is 4, 5, 7, 8, 10, 11, 13, 15 and 17, after the cube is processed, the processing tag value is set to 4; When the corresponding topological configuration of the cube is 2, 3, 6, 9, 12, 14, and 16, after the cube is processed, the processing flag value is set to 3; When the topological configuration of the cube is 1, after the cube is processed, the processing mark value is added by 1.

When the processing tag value of a cube is 4, there is no need for any subsequent processing of the cube. When the growth tag value of the cube is 4, there is no need to put the growth information about the cube into the growth queue. In addition, when the processing label of a cube is greater than 1 and less than 4, only the iso-surface of topological configuration 1 is extracted in the cube.

# 3.Results and discussion

**3.1Ambiguous hole detection result**

For the algorithm proposed in this paper, the primary problem is to solve the ambiguity of topology configuration. Because of the ambiguity, the generated 3D model will have holes in some details, which greatly reduces the overall quality of the model. Aiming at the hole problem, Masala proposed a current simplest solution. It deleted one of the original 15 topological configurations of the MC algorithm, and then added 7 more, and finally set the topological configuration to 21 kinds.

For the inspection of whether the model generated by the 3D reconstruction has holes, this article uses the data used by Masala, which is called MC example by Masala, and its size is $10 \times 10 \times 10$, that is, there are 10 layers of data and each layer is of size $10 \times 10$. Except for the 5th, 6th and 7th layers, the data value of the other layers are all 0. The data of the 5th, 6th and 7th layers are shown in figure 11. The data of the 6th and 7th layers are the same.

```
0 0 0 0 0 0 0 0 0 0          0 0 0 0 0 0 0 0 0 0
0 0 0 0 0 0 0 0 0 0          0 0 0 0 0 0 0 0 0 0
0 0 a 0 0 0 0 0 0 0          0 0 a a 0 0 0 0 0 0
0 0 0 a 0 0 0 0 0 0          0 0 a a a 0 0 0 0 0
0 0 0 0 a 0 0 0 0 0          0 0 0 a a a 0 0 0 0
0 0 0 0 0 a 0 0 0 0          0 0 0 0 a a a 0 0 0
0 0 0 0 0 0 a 0 0 0          0 0 0 0 0 a a a 0 0
0 0 0 0 0 0 0 a 0 0          0 0 0 0 0 0 a a 0 0
0 0 0 0 0 0 0 0 0 0          0 0 0 0 0 0 0 0 0 0
0 0 0 0 0 0 0 0 0 0          0 0 0 0 0 0 0 0 0 0
```

  (a) Layer 5 data          (b) Layer 6 and 7 data

**Fig.11 The data of MC example**

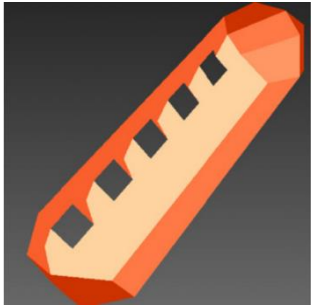 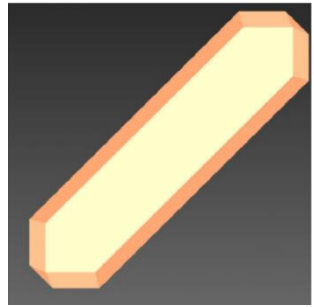 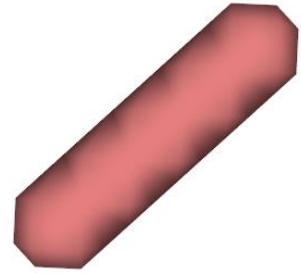

(a)Standard MC algorithm     (b)Masala's        (c)Our

**Fig.12 The 3D reconstruction model of MC example**

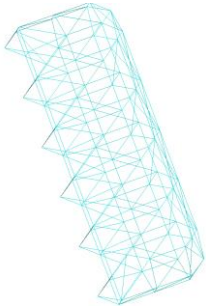 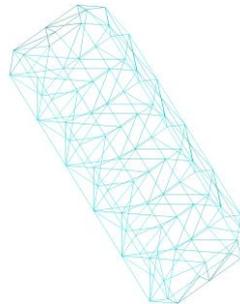 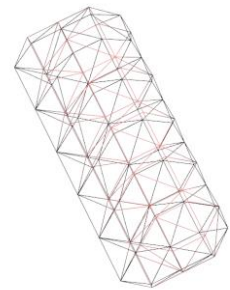

(d)Standard MC algorithm     (e)Masala's        (f)Our

**Fig.13 The 3D reconstruction of grid model of MC example**

  When using the set of data of MC example for reconstruction, the reconstruction results of the standard MC algorithm, Masala's improved MC algorithm and the algorithm used in this article are shown in figure 12. It can be seen from the figure that the reconstruction results of the standard MC algorithm have obvious holes. The reconstruction results of the algorithm used in this article are the same as the improved MC algorithm of Masala, and the generated 3D model has no holes. Figure 13 shows the mesh display results of each model in figure 12. It can be seen that the standard MC algorithm mesh model has obvious jaggedness on the left side. Because of the lack of some triangles on these saw teeth, the final generated model is incomplete and there are some holes. The mesh model corresponding to the improved MC algorithm of Masala is roughly the same as the mesh model corresponding to the algorithm in this paper, and all generated are a closed mesh model.

## 3.2 Quality and time analysis of reconstruction model

The improved MC algorithm of Masala mainly improves the topology configuration of the original MC algorithm, which has the problem of ambiguity. In order to test the performance of our algorithm, two sets of experiments are used to test the reconstruction effect. Table 1 is the environment configuration of the whole experiment.

Table 1 Experimental environment configuration

| Name | Parameter |
| --- | --- |
| Operating system | Win10 64 |
| CPU | i5-8400 |
| Memory | 8GB |
| Development environment | VS2015 |
| platform | VTK |

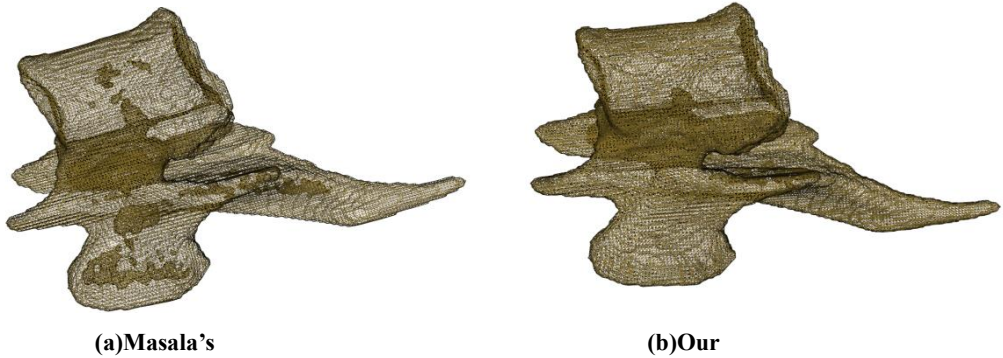

(a)Masala's  (b)Our

Fig.14 3D reconstruction model of the first group of experiment

The first set of experimental data is a set of segmented medical image data, which consists of 145 two-dimensional tomographic images with a size of $125 \times 105$. The 3D reconstruction model of Masala's improved MC algorithm and the algorithm in this paper is shown in figure 14. The seed triangle of the algorithm in this paper is obtained from all the cube voxels in the middle layer. It can be seen from the figure that the two algorithms have the same grid on the overall 3D external contour of the 3D image, but from the inside of the model, the improved MC algorithm of Masala has a lot more internal grids than the model reconstructed by the algorithm in this paper. In practical applications, these internal grids do not participate in the display of the model. These internal grids are mainly caused by incomplete segmentation. In the case that only a complete external 3D contour mesh model is needed, the algorithm in this paper is better than the improved MC algorithm of Masala in the quality of the reconstructed model.

Table 2  3D reconstruction information of the first set of experiment

| Algorithm | Time(s) | Triangular faces |
| --- | --- | --- |
| Masala's | 0.219711 | 79728 |
| Our | 0.212538 | 71472 |

Under the same environment configuration, the time and the number of triangles spent on the first set of data for the 3D reconstruction of the two algorithms are recorded in Table 2. It can be seen that the algorithm in this paper is less than Masala's improved MC algorithm in the number of triangles in the reconstruction model. In terms of reconstruction time, there is little difference between the reconstruction of the two algorithms. The algorithm in this paper is slightly smaller than the improved MC algorithm of Masala.

In the second experiment, 4 sets of data were reconstructed, and the same reconstruction threshold was used for these 4 sets of data. These 4 sets of data are 1#, 2#, 3# and 4# respectively. At the same time, 3D reconstruction was carried out using Masala improved MC algorithm and the algorithm in this paper. The reconstruction result is shown in figure 15, where the seed triangle of the algorithm in` this paper is obtained from all the cube voxels in the middle layer.

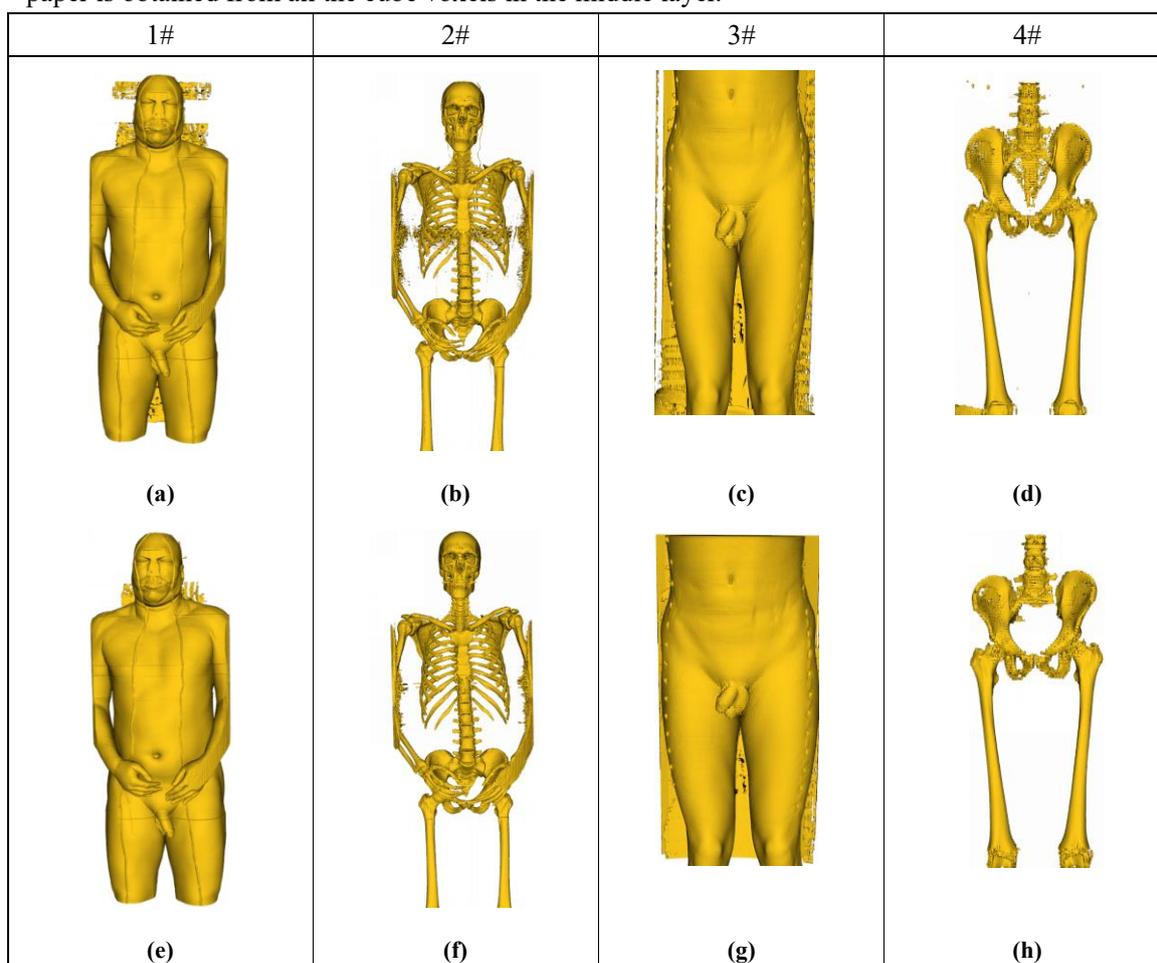

**Fig.15 The 3D reconstruction model of the four sets of data of the second set of experiment**

In figure 15, figures (a), (b), (c) and (d) are 3D models reconstructed by Masala's improved MC algorithm under data 1#, 2#, 3# and 4#, respectively, (e), (f), (g) and (h) are the 3D models reconstructed by the algorithm used in this paper under data 1#, 2#, 3# and 4#. It can be seen that the 3D reconstruction result of Masala improved MC algorithm is similar to that of the algorithm presented in this paper. Figure 16 is a partial enlargement of the chest using the second set of data for the 3D reconstruction. It can be seen that the model of Masala's improved MC algorithm has many scattered small fragments compared with the algorithm in this paper, which is somewhat similar to the noise in the two-dimensional image. These fragments reduce the overall display effect of the model. Therefore, when some fragments that are not connected to the main contour are not needed, the reconstruction quality of this algorithm is better than that of Masala's improved MC algorithm.

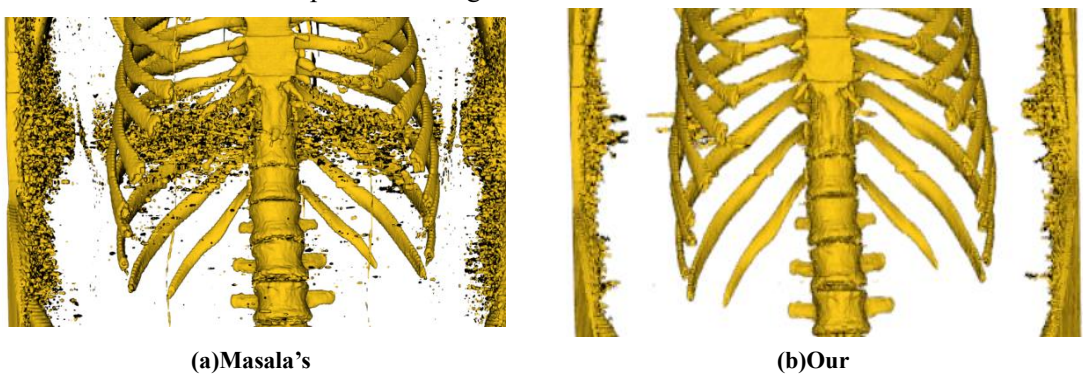

(a)Masala's　　　　　　　　　　　　　　　(b)Our

**Fig.16 Partial enlarged view of the 3D reconstruction model of data No. 2**

**Table 3 Comparison of the 3D reconstruction results of the four sets of data in the second set of experiment**

| Data | Algorithm | time(s) | Triangular faces |
|---|---|---|---|
| 1# | Masala's | 40.1304 | 10706161 |
|  | Our | 36.9829 | 6633879 |
| 2# | Masala's | 33.3153 | 6885207 |
|  | Our | 36.463 | 5973771 |
| 3# | Masala's | 3.9786 | 1519788 |
|  | Our | 3.9027 | 1276479 |
| 4# | Masala's | 3.5484 | 845165 |
|  | Our | 3.4459 | 686160 |

Table 3 shows the time spent by the two algorithms in the reconstruction process of the second set of experiment and the number of triangles in the mesh model. It can be seen that, under the same reconstruction parameters, the number of triangles in the reconstruction model of this algorithm is less than that of Masala's improved MC algorithm. In terms of reconstruction time, the algorithm used in this article takes more time to reconstruct the 2# data. The reconstruction time of the other three sets of data is less than that of Masala's improved MC algorithm.

Through experiments, the improved MC algorithm of Masala is compared with the algorithm adopted in this paper, and it is known that both algorithms can effectively avoid holes in the surface of the 3D reconstruction model. In the algorithm used in this paper, in the 5 sets of data experiments, only one set of experiments takes longer time for 3D reconstruction, while the remaining 4 sets of data take less time. And the number of triangles in the reconstruction model of the algorithm in this paper is less than that of Masala's improved MC algorithm. For the details of the two-dimensional data that do not require 3D reconstruction, the algorithm used in this paper performs better.

**3.3 The influence of seed region selection on reconstruction model**

In order to facilitate the use of this algorithm for 3D reconstruction of medical images, we made a simple 3D reconstruction system. The system can set the selected area of the seed triangle to reconstruct a 3D model connected to the seed triangle. We selected a set of human brain data to test the system. The data consists of 320 two-dimensional images. The results of the 3D reconstruction are shown in figure 17, where figure(a) is a model of the entire data reconstruction using the MC algorithm. The model is the entire skull corresponding to the data, and each bone is scattered in each area. Figure(b) is the reconstruction result of the system when the seed region is $\{(x, y, z)|0 \leq x < 400, 0 \leq y < 400, 0 \leq z < 10\}$, and its corresponding 3D model is the maxilla. Figure(c) is the reconstruction result of the system when the seed selection area is $\{(x, y, z)|200 \leq x < 300, 400 \leq y < 478, 270 \leq z < 318\}$, and its corresponding 3D model is the mandible. It can be seen that when the system selects different seed regions, the reconstructed model is also different. Based on this reconstruction effect, for multiple scattered tissues, the algorithm in this paper can be used to set the seed area to perform 3D reconstruction of the designated part of the image.

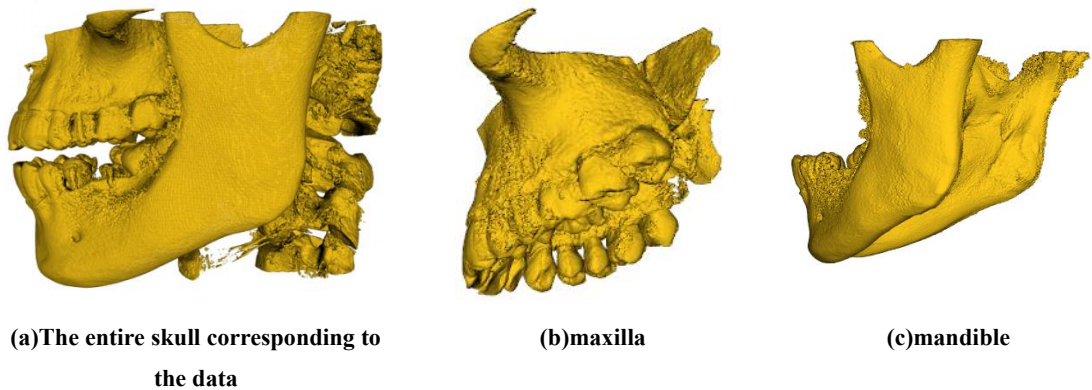

(a)The entire skull corresponding to the data  (b)maxilla  (c)mandible

**Fig.17 The 3D reconstruction results of seeds selected from different regions in the human brain**

## 4.Conclusion

This paper proposes a Marching Cube algorithm based on edge growth for 3D reconstruction. First of all, the 3D model reconstructed by this algorithm does not have holes caused by ambiguity. Secondly, from the perspective of overall reconstruction quality, it is consistent with the appearance of the model reconstructed by Masala's improved MC algorithm, but when only the main contour of the model is needed, the reconstruction effect of this algorithm is better. In terms of reconstruction accuracy, this algorithm and MC and Masala's improved MC algorithm are based on cube voxel interpolation for 3D reconstruction, so their reconstruction accuracy is the same. In the application of this algorithm, when the tissues in the data are multiple scattered parts, the 3D contour of the specified part can be extracted by setting the region of the selected seed.

For the algorithm in this article, its main limitations are as follows: The algorithm uses a queue to store the growth side information. Compared with other types of algorithms, it requires more memory space and is not suitable for large-scale reconstruction on a small machine. This algorithm generates a mesh model based on seeds, and is not suitable for reconstructing some scattered multiple objects such as broken bones. It needs to artificially set seeds for each scattered part, or expand the area of seed selection.

In the subsequent development, multithreading can be used to accelerate the reconstruction speed of the algorithm. It can also be combined with volume rendering. First, determine the area to be reconstructed according to the display results of volume rendering, then set the seed area and reconstruction threshold, and use the surface rendering algorithm in this paper to accurately reconstruct the tissue of the specified part.